\documentclass[12pt]{article}
\usepackage{graphicx,lineno}
\newcounter{abceqn}
\def\abclabel#1{\@bsphack\if@filesw{\let\thepage\relax%
	\def\protect{\noexpand\noexpand\noexpand}%
\xdef\@gtempa{\write\@auxout{\string%
\newlabel{#1}{{\arabic{abceqn}}{\thepage}}}}}\@gtempa%
\if@nobreak\ifvmode\nobreak\fi\fi\fi\@esphack}
\def\abceqnbeg{\setcounter{abceqn}%
		{\value{equation}}\addtocounter{abceqn}{1}%
		\setcounter{equation}{0}%
		\def\theabceqn{\arabic{abceqn}\alph{equation}}%
		\def\theequation{\theabceqn}}
\def\abceqnend{\def\theequation{\arabic{equation}}%
		\setcounter{equation}{\value{abceqn}}}
		       {\end{eqnarray}\abceqnend\newline}
\renewcommand{\baselinestretch}{1}
\newcommand{\eqn}[1]{Eq.(\ref{#1})}
\newcommand{\ovln}[1]{\overline{{#1}}^{obs}}
\newcommand{\F}{F_{out}}
\newcommand{\wheps}{\widehat{\epsilon}}
\newcommand{\Ts}{T_s}
\newcommand{\Teq}{\Delta \Ts^{eq}(X/X_o)}
\newcommand{\TEQ}{\delta T_{f}}

\newcommand{\Tcs}{\lambda_{2X_o}^{eq}}
\newcommand{\Td}{\lambda_{2X_o}^{obs}}
\newcommand{\dTdt}{d(\ovln{\Delta \Ts})/dt}
\newcommand{\fb}{\phi_{\mbox{\small fb}}}
\newcommand{\avg}[1]{\overline{#1}}
\renewcommand{\gamma}{\tau_*}
\newcommand{\tauT}{\tau^{obs}_T}
\newcommand{\tauS}{\tau_{*}}
\begin{document}
\title{The Earth's Climate Sensitivity \\ and Thermal Inertia}
\author{
B.~S.~H. Royce$^*$ and 
S.~H.~Lam\thanks{Professors emeritus. Corresponding author email: bshroyce@princeton.edu.}\\
Department of Mechanical and Aerospace Engineering\\
                Princeton University\\ Princeton, NJ 08544, U.~S.~A.}
 \date{February 15, 2011\\ \emph{Revised} \today}
\maketitle
\pagestyle{myheadings}
\markright{\today \hfil Royce and Lam  \hfil}
\newpage
\abstract{The Earth's \emph{equilibrium climate sensitivity}  has received much attention because of its  relevance and importance for global warming policymaking. This paper focuses on the Earth's  \emph{thermal inertia time scale}     which has  received relatively little attention.  The difference between the 
observed \emph{transient climate sensitivity} and the equilibrium climate sensitivity  is shown to be proportional to the thermal inertia time scale, and the numerical value of the proportionality factor is determined using recent observational data.
 Many useful policymaking insights can be extracted from the resulting empirical quantitative relation.

\newpage

\section{Introduction}

For millennia prior to the industrial revolution, the global surface temperature of the Earth, $T_s(t)$, fluctuated about a constant average value $T_{s,o}$ ($\approx 288 ^\circ K$), while the  atmospheric carbon dioxide concentration $X(t)$ fluctuated about a constant average value $X_o$ ($\approx 285$ ppmv).   Subsequent to 1850, the value of $X(t)$ has increased significantly above $X_o$, and the  observed average global surface temperature $T_{s}(t)$ has also increased.  Since carbon dioxide is a major long resident time greenhouse gas, it is reasonable to associate the observed  increase of $T_{s}(t)$ to the
increase of $X(t)$ and other  atmospheric greenhouse gases.
A useful quantitative measure of atmospheric carbon dioxide driven global warming  is the estimated amount of average warming $\Delta T_s$ (above the pre-industrial revolution value $T_{s,o}$)  when $X/X_o$ is held at some constant value until the Earth system reaches   \emph{steady state equilibrium}. This \emph{equilibrium global surface temperature rise} is denoted by
$\Teq$.
The \emph{Intergovernmental Panel on Climate Change} (IPCC) (Solomon  \emph{et al.}, 2007; Houghton  \emph{et al.}, 2001, 1995, 1990)  
recommends the following relationship for $\Teq$: 
\begin{equation}
\Teq \equiv \frac{\Tcs}{\ln 2}\ln\left( \frac{X}{X_o}\right). \label{eqcs}
\end{equation}
While $\Teq$ depends logarithmically on $X/X_o$, it is merely proportional to $X/X_o-1$ for $X/X_o$ close to unity.

The constant parameter $\Tcs$ in \eqn{eqcs}   is called the   \emph{equilibrium climate sensitivity} of the Earth for carbon dioxide.   It  is the  average global surface temperature rise that results when $X$ is held constant at $2X_o$ until steady state  is reached.
The numerical value of $\Tcs$ is crucial to global warming policymaking. The IPCC suggested
$2^\circ C \leq \Tcs \leq 4.5^\circ C$ as its  ``likely range,'' and $\Tcs \approx 3^\circ C$ as its ``best estimate'' value. These suggestions were based largely on assessments of data generated by the best available computer
climate models (Houghton  \emph{et al.} 1990, Box 10.2; also Andronova and Schlesinger, 2001). 
After acknowledging that
the numerical value of $\Tcs$ ``... is a key uncertainty ...'', the IPCC nevertheless said ``... values substantially higher than $4.5^\circ C$ cannot be excluded, but agreement of models with observations is not as good for those values.'' 
(Solomon  \emph{et al.},  2007, \S 2.3, Chapter 10, p.798. See also Houghton  \emph{et al.}, 1990, \S8.6, \S9.6, SPM). 
This statement implies that 
the \emph{probability distribution function} (PDF) of $\Tcs$ might have a ``fat-tail,''   an issue of concern in the policy community
(Weitzman, 2009).

\section{Historical observational data}

Fig.\ref{Hadley} plots  observational average global surface temperature $\Delta \Ts$  data versus $\ln (X/X_o)$ between 1850-2012 (Brohan  \emph{et al.}, 2006; Enting  \emph{et al.} 1994; Morice \emph{et.~al.} 2012). The $X(t)$ data is a relatively smoothly rising function of time (See Fig.~1 in Hansen \emph{et al.}, 2005). The data for $\ln(X/X_o)<0.13$ were taken
before 1970. The scatter of the pre-1970 data is significantly larger than that of the post-1970 data (see also Fig.~4A in Hansen \emph{et al.}, 2010).

Inspection of Fig.\ref{Hadley}  suggests the following formula for the post-1970 observed data:
\begin{equation}
\ovln{\Delta T_s}(t)=\frac{\Td}{\ln 2}\ln\left( \frac{X(t)}{X_o}\right), \label{tcr}
\end{equation}
where the overline notation denotes some ``best-fit'' representation of the data over some ``sufficiently long'' time interval, and $\Td$,
the transient climate sensitivity (Held, Winton, Takahashi \emph{et.~al.}, 2010),  is a constant to be empirically determined by curve-fitting. 
From a least squares fit to the post-1970 data,  one obtains $\Td\approx 2.3^\circ C$ (R=0.88). In Fig.\ref{Hadley}, the bold line is this best-fit $\ovln{\Delta T_s}(t)$ line,\footnote{Note that the bold line in Fig.~\ref{Hadley} misses the origin by a small amount because it is the best-fit line only for the post-1970 data points. If all the pre-1970  $\Delta \Ts$ data in Fig.~\ref{Hadley} were also included, the resulting best-fit line would pass through the origin with $\Td\approx 1.9^\circ C$ (correlation coefficient $R=0.86$).} 
while the dashed line  is  $\Delta T_s^{eq}(t)$  plotted using \eqn{eqcs} with $\Tcs=4.5^\circ C$---which is at the upper limit of the IPCC likely range.  
The divergence between these two lines is clearly shown with observed values following the 2.3$^\circ$C line.
A good correlation is necessary for  the cause-and-effect connection between the observed warming and atmospheric carbon dioxide, but of course it is {not} sufficient to ``prove'' the connection.

The value of $\Td$ is the amount of average temperature rise \emph{at the moment }when a \emph{steadily rising}   $X(t)$ reaches twice the pre-industrial revolution level. Obviously, the value of $\Td$ depends on the  rate    of increase in $X(t)$. When $X(t)$ increases at $1\%$ per year, the value of $\Td$ is called the  \emph{transient climate response} (TCR).  (See Fig.~10.25 of Houghton  \emph{et al.}, 1990). The post-1970 $X(t)$ observational data is indeed steadily rising at about $0.5\%$ per year and  if this rate is small enough, then the observed value of $\Td$ should be a good approximation to $\Tcs$.

\subsection{Rising rate of ${\Delta T_s}(t)$}

Fig.~\ref{Slope} is a plot of the post-1970  observational $\Delta T_s(t)$ data versus time  (Brohan  \emph{et al.}, 2006; see also Fig.~2 in Hansen \emph{et al.}, 2005).
It is seen that this data exhibits considerable scatter that can be attributed to natural random events such as El Ni$\tilde{n}$os, solar irradiance variations, albedo changes caused by major volcanic eruptions, etc.
A more meaningful value of $d\Delta T_s/dt$ would be the slope of some \emph{appropriately chosen} best-fit  $\ovln{\Delta T_s}(t)$ versus time line.  

A simple exponential is chosen to represent the post-1970 observational data, that is, 
$\ovln{\Delta T_s}(t)$ is assumed to be described by the following differential equation:
\begin{equation}
\frac{d\ovln{\Delta T_s}}{dt}= \frac{\ovln{\Delta T_s}}{ \tauT},   \label{tauTs}
\end{equation}
where the constant $\tauT$ is to be empirically determined by curve-fitting.
Using least-squares, the value $\tauT$ of the post-1970 $\Delta T_s(t)$ data is found to be approximately $32$ years ($R=0.90$). The resulting
best-fit $\ovln{\Delta T_s}(t)$ is the
bold solid line in Fig.~\ref{Slope}.  

Of course,  other time dependencies  could have been chosen instead of \eqn{tauTs} to curve-fit the post-1970 $\Delta T_s(t)$ data. 
The exponential time dependence was chosen because it has the correct \emph{qualitative} behavior.

\section{Energy balance of the Earth}

The difference between the transient and the equilibrium $\Delta T_s$ responses  is attributed to the finite thermal inertia of the Earth system (Hansen \emph{et al.}, 2005).
The Earth receives radiant energy from the Sun, and emits radiant energy back into space. The mismatch of these two fluxes  changes  the thermal energy content of the Earth system.
Using a control volume enclosing the Earth (with its boundary surface \emph{above} the atmosphere), the Earth's energy balance equation is:
\begin{equation}
\frac{d E}{dt}=F_{in}(t)-\F(X,T_s,\ldots),   \label{ebalance}
\end{equation}
where $E$ is the stored thermal energy content per unit area of the Earth (Joules per unit area), and $F_{in}(t)$ and $\F(X,T_s,\ldots)$ are, respectively, the incoming radiant energy flux  from the Sun and the outgoing radiant energy flux 
away from the Earth---both fluxes (Joules per year per unit area)  evaluated \emph{at the top of the atmosphere}.   The left hand side represents the thermal inertia of the Earth system.

\subsection{Formulation and assumptions}

For millennia prior to the industrial revolution, the Earth's $T_s(t)$ 
fluctuated about  a constant steady state value $T_{s,o}$ of circa $288^\circ K$. Current warming of the Earth is assumed to be caused mainly by the greenhouse effects of  increasing atmospheric carbon dioxide $X(t)$.     The amount of equilibrium warming, \eqn{eqcs},   contains a single time-independent parameter $\Tcs$, and 
it is desired to extract as much information as possible about $\Tcs$ from the post 1970 observational data without getting too deeply involved with the detailed  physics of the atmosphere and the oceans (Bierbaum  \emph{et al.}, 2003, Gregory  \emph{et al.}, 2002).

To this end, two simplifying assumptions are adopted to formulate the problem: 
\begin{enumerate}
\item $E(T_s)$ depends only on $T_s$. This is called the \emph{single energy reservoir} assumption
which can be justified when all faster thermal reservoirs have equilibrated with each other so that  a single slower  reservoir dominates the system.  (See Held \emph{et.~al.} (2010), Stouffer (2010), and Socolow and Lam (2007) for the details of the mathematics.)
\item $\F=\F(X,T_s)$, i.e. $\F$ depends only on $X$ and  $T_s$, and nothing else. This assumes that atmospheric carbon dioxide, which has a long atmospheric life time, is the dominant  cause of the global warming problem. It is known that there are other greenhouses gases in the atmosphere (e.g. methane, which has a much shorter atmospheric residence time than carbon dioxide). {Wigley, Jones and Raper (1997; see its Fig. 5) suggested that atmospheric aerosols can also play a  major role in the energy balance equation, contributing negative radiative forcing by reflecting incoming sunlight. \eqn{eqcs} does not account for such effects.} 
\end{enumerate}
Atmospheric water vapor and other feedback effects will be dealt with in \S \ref{EofTs}.
The role played by the thermal inertia term in \eqn{ebalance} will be explored.

\subsection{Linearized  response to perturbations}

When  $F_{in}(t)$ and $X(t)$  are perturbed   from their pre-industrial revolution steady state values
$F_{in,o}$ and $X_o$ by $\delta F_{in}(t,\ldots)$ and $\delta X(t)$,   the linearized equation governing the response of the surface temperature $\Delta T_s$, derived from \eqn{ebalance}, is:
\begin{equation}
\frac{d E}{d T_s}\frac{d \Delta T_s}{dt}=\delta F_{net}(t,X,\ldots)-\left( \frac{\partial \F}{\partial T_s}\right)_{X} \Delta T_s, \label{two}
\end{equation}
where  $d E/d T_s$ is an \emph{effective specific heat}  per unit area of the Earth,   and $\delta F_{net}(t, X,\ldots)$ is the net incremental amount of  \emph{radiative forcing} per unit area of the Earth:
\begin{equation}
\delta F_{net}(t,X,\ldots)\equiv \delta F_{in}(t,\ldots)-\left(\frac{\partial \F}{\partial X}\right)_{T_s}\delta X(t). \label{Fnet}
\end{equation}
The last term in \eqn{two} represents the incremental amount of energy being radiated away from a warmer Earth.
On the right hand side of \eqn{Fnet}, the first term $\delta F_{in}(t,\ldots)$ 
accounts for all the natural random perturbations such as variations of solar irradiance, albedo changes caused by volcanic eruptions, etc., while  the second term accounts for the \emph{direct} greenhouse effect due to the increase of atmospheric carbon dioxide.

Assuming $(\partial \F/\partial T_s)_{X_o}\neq 0$ and dividing \eqn{two} through by it, one obtains (North, Cahalan and Coakley, 1981; Schwartz, 2007, 2008. See also Held, Winton, Takahashi \emph{et al.} 2010):
\begin{equation}
\tauS \frac{d\Delta T_s}{dt}=\TEQ(t, X,\ldots)-\Delta T_s,\label{unst}
\end{equation}
where  
\abceqnbeg
\begin{eqnarray}
\tauS &\equiv& \frac{ \frac{d E}{d T_s} }{ \left(\frac{\partial \F}{\partial T_s}\right)_{X}}, \label{fiveA}\\
\TEQ(t, X,\ldots) &\equiv&\frac{\delta F_{net}(t, X,\ldots)}{\left(\frac{\partial \F}{\partial T_s}\right)_{X}}.\label{dTdF}
\end{eqnarray}
\abceqnend
Under the single energy reservoir assumption, $\tauS$ is formally a constant. However,  it has been shown that  for a multi-reservoir system $\tauS$ could be time-dependent (see Appendix B of Socolow and Lam, 2007).

Once $\tauS(t)$ is somehow specified, \eqn{unst} can be used to compute the $\Delta T_s(t)$ 
response to any \emph{given} forcing function $\TEQ(t,X(t),\ldots)$ of interest.
Since \eqn{unst} is linear, the response to any additive contribution to $\TEQ(t,X(t),\ldots)$  can be separately studied
(e.g. methane, sulfur aerosols, etc.).

\subsection{Time-averaging} 

The actual forcing function $\TEQ(t,X,\ldots)$ on the Earth system  contains all the unavoidable natural random disturbances.
Thus the observed $\Delta T_s(t)$ data must also contain random components.
Formally,  zero-mean randomness can always be removed in such data by taking a running time-average over some sufficiently long time interval. Applying running time-averaging to  \eqn{unst} and denoting all the time-averaged entities using the overline notation, gives:
\begin{equation}
\tau \frac{d\avg{\Delta T_s}}{dt}=\avg{\TEQ(t, X,\ldots)}-\avg{\Delta T_s}. \label{avgunst}
\end{equation}
where a new time scale parameter $\tau$ is formally introduced to replace $\tauS(t)$. The subtle distinction between
$\tau$ and $\tauS$ will be discussed in \S \ref{five}.

\eqn{avgunst} governs the time-averaged response  of the Earth's system  $\avg{\Delta T_s}(t)$ as driven by a \emph{completely general}  time-averaged forcing function $\avg{\TEQ(t, X,\ldots)}$.   The parameter $\tau$ is as yet unknown.

\section{The use of $\ovln{\Delta T_s}(t)$} \label{usedata}

It is assumed that the time-averaged observational data, $ \ovln{\Delta T_s}(t) \approx \avg{\Delta T_s}$, honors \eqn{avgunst}. Using \eqn{tauTs} to eliminate $d\ovln{\Delta T_s}/dt$ from  \eqn{avgunst}, gives:
\begin{equation}
\avg{\TEQ(t, X,\ldots)}\approx (1+\frac{\tau}{\tauT})\ \ovln{\Delta T_s}(t). \label{forcingfn}
\end{equation}
This $\avg{\TEQ(t, X,\ldots)}$ forcing function is \emph{empirically consistent} with the actual post-1970 $\ovln{\Delta T_s}(t)$ observed data for any positive $\tau$. The amount of  unrealized global warming ``in the pipeline''   (Hansen \emph{et al.}, 2005)   is thus $(\tau/\tauT)\ovln{\Delta T_s}(t)$.

\emph{Note that  no detailed physics was required   in the derivation of
\eqn{forcingfn}}. The crucial enabling step was the successful curve-fitting of the post-1970 $\Delta T_s(t)$ observational data to a simple exponential. 

Using \eqn{tcr} for $\ovln{\Delta T_s}(t)$, one obtains:
\begin{equation}
\avg{\TEQ(t, X,\ldots)}\approx (1+\frac{\tau}{\tauT})\frac{\Td}{\ln 2}\ln(\frac{X}{X_o}).\label{foura}
\end{equation}
Comparing \eqn{foura} to \eqn{eqcs},  a simple formula for the equilibrium climate sensitivity $\Tcs$ is obtained:
\begin{equation}
\fbox{$\Tcs\approx (1+\frac{\tau}{\tauT})\Td$.} \label{sensit}
\end{equation}
The right hand side of \eqn{sensit} contains  the unknown $\tau$ parameter. The values of the other two parameters, $\Td\approx 2.3  ^\circ C$ and $\tauT\approx 32$ years, have been  determined from  the post-1970 observational data
using least-squares curve-fitting. 

Interesting questions are: is the order of magnitude of $\tau$ bigger or smaller than 32 years?
Can $\tauS$ be determined empirically by examining the actual $\Delta T_s(t)$ data? Can the order of magnitude of $\tau$ be estimated from such empirical values of $\tauS$?

\section{Observational estimates of $\tau$}\label{five}

The Earth system is expected to exhibit a number of different thermal response time scales depending on the mode of excitation.
The governing equation for $\Delta T_s(t)$ is \eqn{unst}, and the role played by  $\tauS$
on the $\Delta T_s(t)$ response to either a periodic or a Dirac-delta forcing function is well known.
Since Earth's orbit around the sun has a small eccentricity, the solar radiation flux arriving on the Earth has a significant annual periodic variation ($\delta F_{net}(t)/F_{in,o}\approx 0.03 \sin(2\pi t)$). Douglass, Blackman and Knox 
(2004) analyzed the relevant observational data and found that the  value of $\tauS$ inferred from the phase shift of the data is less than one year (i.e. $2\pi \tauS=O(1)$), while that inferred from the amplitude data is significantly larger than one year (i.e. $2\pi \tauS>>1$).  They favored the small $\tauS$ inference from the phase shift data, and  explained the discrepancy with the amplitude inference by a large \emph{negative} feedback factor. 
The 1991 Pinatubo eruption is known to have affected the world's weather for several years. The commonly accepted explanation is
that volcanic aerosols released into the atmosphere  reduced incoming solar radiation by reflection
(i.e. negative radiative forcing), and the residence time of such aerosols in the atmosphere is several years.
Thus the 1991 Pinatubo eruption can be approximated by a Dirac-delta forcing function.
Douglass and Knox  (2005, 2006) analyzed the available observational data, and concluded that the inferred value of $\tauS$ was less than one year.

The derivations of \eqn{unst} and \eqn{avgunst}  both adopt the \emph{single energy reservoir assumption}.  The real Earth system obviously has more than one thermal energy reservoir, and all the participating reservoirs are expected play a role.  
 Thus the empirically determined $\tauS$'s discussed above were the time scales of  the  fast reservoirs which responded to the high frequency forcing. The value of $\tau$ of interest in this paper is the time scale of the slower reservoir which responded to the radiative forcing in the post-1970 time period. The bottom line is that $\tau$ and $\tauS$ are not the same numerically. 
The values of the $\tauS$'s  can only provide a lower limit to the order of magnitude of $\tau$.

\section{The physics of  $\tau$} \label{EofTs}

The difficulties of modeling and calculating $\tau$ using first principles can be appreciated by assuming $\tau\approx \tauS$ and attempting   it with \eqn{fiveA}.

The numerator
$(dE/d\Ts)_o$  on the right hand side of \eqn{fiveA}  is the effective specific heat per unit surface area of the Earth. Assuming  the oceans to be the Earth's sole
energy reservoir, one can represent $(dE/d\Ts)_o$ by the product of the
 per unit volume specific heat of water ($4.2 \times 10^6 Joule/m^3\mbox{\--\-}^\circ C$)
and an \emph{effective energy storage depth} $H$ of the oceans:
\begin{eqnarray}
\left(\frac{dE}{dT_s}\right)_o&=&H\times 4.2 \times 10^6 \quad (Joule/m^2\mbox{\--\-}^\circ C) \nonumber\\
&=& 0.13  H \quad (W\mbox{\--\-}year/m^2\mbox{-}^\circ C),  \label{Honethree}
\end{eqnarray}
where $H$ is in meters.  \emph{All the difficult physics of mass and energy transport in the oceans is contained in $H$}.

The denominator $(\partial F_{out}/\partial T_s)_{X_o}$ on the right hand side of \eqn{fiveA}  is determined by 
the detailed physics of radiative energy transport in the atmosphere. 
The outgoing radiation flux $F_{out}(T_s,X)$  at the top of the atmosphere is the sum of (i) the black body radiation emitted at the Earth's surface  and (ii) the thermal radiation emission of
the atmosphere itself, both \emph{duly attenuated} by atmospheric absorption as they emerge from the atmosphere.
One may formally represent $F_{out}(T_s,X)$ by:
\begin{equation}
F_{out}(T_s,X)=\wheps(T_s,X,\ldots) \sigma T_s^4\quad (W/m^2),\label{SB}
\end{equation}
where  $\wheps(T_s,X,\ldots)$ is the \emph{effective grey body emissivity} of the Earth at the top of the atmosphere looking downward, and $\sigma$ is the  Stefan-Boltzmann constant. \emph{All the difficult physics inside
the atmosphere (i.e. the $T_s$ dependence of atmospheric water vapor, clouds, glaciers, conditions of the troposphere, etc.) is   
contained in the $\wheps(T_s,X,\ldots)$ factor.}

Taking the partial derivative of \eqn{SB} with respect to $T_s$, \emph{assuming} $\wheps(T_s, X)$, and rearranging, one obtains:
\begin{equation}
\left(\frac{\partial F_{out}}{\partial T_s}\right)_{X_o}=(4-\fb)\frac{F_{out,o}}{T_{s,o}},\label{seven}
\end{equation}
where $\fb$ is:
\begin{equation}
\fb\equiv-\left(\frac{\partial \ln \wheps}{\partial \ln T_s}\right)_{X_o}.
\end{equation}
This dimensionless $\fb$ represents the net  ``feedback factor''  of the Earth's atmosphere. 
The conventional wisdom is that $\fb$ is positive, and that its value is dominated by water vapor feedback. 
Water vapor alone cannot be responsible for climate change because its atmospheric concentration is controlled by the atmospheric temperature distribution.  Atmospheric Carbon dioxide, however, modifies this distribution via the greenhouse effect (Stevens and Bony, 2013).
\emph{First principle estimates of $\fb$  have large uncertainties because the $T_s$ dependence of
$\wheps(T_s,X,\ldots)$ is very difficult to model.}
Note also that $(4-\fb)$ is the ``effective $T_s$ exponent'' of \eqn{SB}. Since \eqn{unst} has been stable for millennia,   
$(4-\fb)$ must be positive to be consistent with this historical fact. 

The values of $dE/dt \ (W/m^2)$ and $d\Delta T_s/dt \  (^\circ C/year)$ are related by $(dE/dT_s)_o$ which is represented by \eqn{Honethree}:
\begin{equation}
\frac{dE}{dt}(W/m^2)=0.13H \times \frac{d\Delta T_s}{dt}(^\circ C/year). \label{Heqn}
\end{equation}
Recently, Schwartz \emph{ \emph{et al.}} (2010) carefully reviewed the literature on the available observational data of upper ocean heat content (Gregory  \emph{et al.}, 2002; Schwartz, 2007; Willis  \emph{et al.}, 2004; Gouretski and Koltermann, 2007; Wijffels  \emph{et al.}, 2008), and recommended $\overline{dE/dt}^{obs} \approx 0.37 \ W/m^2$---which is significantly smaller than the value ($ 0.60 \ W/m^2$) previously recommended by Hansen \emph{et al.} (2005).
Using these two values  in \eqn{Heqn} along with a rough estimate for $\dTdt$ from Fig.~\ref{Slope}, one obtains
$H\approx 140-230$ meters. The order of magnitude of storage depth, $H$, values in this range is very reasonable (Ross, 1982).

Using the above  (and $F_{out,o}\approx 250\ W/m^2, T_{s,o}\approx 288 ^\circ K, H\approx 190$ meters, $\tau \approx \tauS$)  in \eqn{fiveA}, one obtains:
\begin{equation}
\fbox{$\tau=\frac{0.13 H }{4-\fb}\frac{T_{s,o}}{F_{out,o}}\approx \frac{28}{4-\fb}\quad years.$} \label{taudTdt2}
\end{equation}
 When one only knows the order of magnitude of $\tau$,  \eqn{taudTdt2} can be used to provide a credible estimate of $\fb$. In contrast,  \eqn{taudTdt2} should \emph{not} be used to estimate $\tau$
if  $\fb$  is expected to be an uncertain number close to $4$.

\section{Summary and concluding remarks}

This paper has been updated using observational data published since 2008 (Morice, C. P. \emph{et.~al.}, 2012). This new data is completely consistent with that used previously as can be seen from Fig.1, 2.
\begin{itemize}
\item The data shown in Figs. 1 and 2 has been fit using least squares to yield values of the transient climate sensitivity ($\approx 2.3^\circ C$) and the thermal inertia time scale associated with global warming ($\approx 32$ years).
\item[] The single energy reservoir assumption is crucial to the derivation of Eq.(12) and Eq.(18) which are the main results of the paper. No detailed physics was required in the derivation of Eq.(12) which relates the Earth's equilibrium climate sensitivity $\Tcs$ to its thermal inertial time scale.
\item \eqn{sensit} also says  that the lower limit of $\Tcs$ is unlikely to be less than $2.3 ^\circ C$ since the factor $\tau/\tauT$ must be positive.  The IPCC best estimate of $\Tcs\approx 3^\circ C$ would need $\tau\approx (3-2.3)\times (31/2.3)\approx 9.4$ years. A fat-tailed $\Tcs$  beyond the IPCC  $4.5 ^\circ C$ upper limit would need $\tau  \ge 32$ years.  

\item \eqn{taudTdt2} relates the thermal inertia time scale $\tau$ to the atmospheric feedback factor $\fb$ and   a parameter $H$ which is an effective energy storage depth of the Eath's oceans. 
Using available  observational data of the upper ocean heat content gives an estimated value of $H$  between 140 and 230 meters.
 \end{itemize}

Note that $\Delta T_s$ will continue to rise even after $X/X_o$ has been successfully stabilized at some constant value.
For example, if $\tau\approx 32$ years, then it would take more than six decades after $X/X_o$ is stabilized in order for
$\Delta T_s$ to eventually reach $90\%$ of its full equilibrium rise. When $\tau$ is some multi-decadal number, the Earth's response  to changes of radiative forcing is very sluggish. This sluggishness has important policy implications.

What is the probability that $\tau$ might be a multi-decadal or even multi-centurial number? It is not  obvious that this question could ever be answered \emph{objectively}. The meaningfulness of probability distribution functions constructed by polling modeling data generated by computers is still subject to debate.  
At the present time, the available observational data do not support \emph{subjective} assignments of multi-decadal  $\tau$'s.

\subsubsection*{Acknowledgment}
The authors acknowledge many valuable discussions  with Dr.~Carleton H.~Seager of Sandia National Laboratory, Dr. Douglas MacMynowski of the Califorinia Institute of Technology,  and Professor Robert Socolow of Princeton University during the course of this work.
 
\newpage
\section*{References}

\begin{description}
\item[ ] Andronova, N.~G. and Schlesinger, M.~E., 2001: Objective estimation of the probability distribution function of climate sensitivity,
\emph{Journal of Geophysical Research}, {\bf 106}, No.~D19, pp.~22605-22611.
\item[] Bierbaum, R.~M., Prather, M.~J., Rasmusson, E.~M., and Weaver, A.~J., 2003: Estimating Climate Sensitivity: Report of a Workshop, National Research Council, The National Academies Press. ISBN 0-309-09056-3.
\item[] Brohan, P., Kennedy, J,~J., Harris, I., Tett, S.~F.~B., and Jones P.~D., 2006: {Uncertainty estimates in regional and global observed temperature changes: a new dataset from 1850}. \emph{J. Geophys. Res}. {\bf 111}, D12106. 
\item[] Douglass, D.~H., Blackman, E. ~G., and Knox, R.~S., 2004: Temperature response of Earth to the annual solar irradiance cycle,
\emph{Phys. Lett. A}., 323, pp.~315-322. See also its corrigendum, \emph{Phys. Lett. A}., 325, pp.~175-176.
\item[] Douglass, D.~H., Knox, R.~S., 2005:  {Climate Forcing by the volcanic eruption of Mount Pinatubo},
\emph{Geophysical Research Letters}, {\bf 32}, L05710. 
\item[] Douglass, D.~H., Knox, R.~S., Pearson, B.~D., and Clark Jr. A., 2006: Thermocline flux exchange during the Pinatubo event,
\emph{Geophysical Research Letters}, {\bf  33}, L19711.
\item[] Enting, I.~G., Wigley, T.~M.~L., and Heimann, M., 1994: Future emissions and concentrations of carbon dioxide, key ocean/atmosphere/land analysis, CSIRO Division of Atmospheric Research Technical Paper No.~31, CSIRO, Australia (updated 2001).
\item[] Frank, P., Uncertainties in the Global Average Surface Air Temperature Index: A representative lower limit. \emph{Energy and Environment}, \textbf{21}, 8 (2010).
\item[] Gouretski, V., and Koltermann, K.~P., 2007: How much is the ocean really warming? \emph{Geophys. Res. Lett.}, {\bf  34}, L01610.
\item[] Gregory, J.~M.~R., Stouffer, R.~J., Raper, C.~B., Stott, P.~A., and Rayner, N.~A., 2002: An observationally based estimate of the climate sensitivity, \emph{J. Climate}, {\bf 15}, (22), pp.~ 3117-3121.
\item[] Hansen, J., Nazarenko, L., Ruedy, R., Sato, M., Willis, J., Denio, A.~D., Koch, D., Lacis, A., Lo, K., Menon, S., Navakow, T., Perlwitz, J., Russell, G., Schmidt, G.~A., and Tausnev, N., 2005: Earth's energy imbalance: confirmation and implications, \emph{Science}, {\bf 308}, pp.~1431-1435.
\item[] Hansen, J., Ruedy, R., Sato, M., and Lo, K., 2010: Global surface temperature change, 
\emph{Rev.~Geophysics}, {\bf 48}, RG4004, doi: 10.1029/2010RG000545.
\item[] Held, I.~M., Winton, M., Takahashi, K., Delworth, T., Zeng, F., and Vallis, G.~K., 2010: Probing the fast and slow components of global warming by returning abruptly to preindustrial forcing, \emph{J. Climate}, {\bf 23},  9, pp.~2418-2427.
\item[]  Houghton, J.~T., Ding, Y., Griggs, D.~J., Noguer, M., van der Linden, P.~J., Dai, X., Maskell, K., Johnson, C.~A., (eds). 2001:   \emph{Climate Change 2001 (TAR): The Scientific Basis} by WG I, Cambridge University Press.
\item[]
Houghton, J.~T., Meira Filho, L.~G., Callander, B.~A., Harris, N., Kattenberg, A., Maskell, K., (eds). 1995: \emph{Climate Change 1995,  (SAR):  The Scientific Basis} by WG I,
Cambridge University Press.
\item[]
Houghton, J.~T., Jenkins, G.~J., Ephraums, J.~J., (eds). 1990: Climate Change 1990, (FAR): \emph{The IPCC Scientific Assessment} by WG I, Cambridge University Press.
\item[] Morice, C.~P., Kennedy, J.~J., Rayner, N.~A., and Jones, P.~D., 2012: Quantifying uncertainties in global and regional temperature change using an ensemble of observational estimates. The HadCRUT4 dataset, \emph{J. Geophys.Res.}, 117, D08101;
 \newline doi: 10.1029/2011JD17187.
\item[] Ross, D.~A., 1982: \emph{Introduction to Oceanography}, 3rd Edition. Prentice-Hall.
\item[] North, G.~R., Cahalan, R.~F., and Coakley Jr., J.~A., 1981: Energy balance climate models, \emph{Rev. Geophys}., {\bf 19}, 1, pp.~91-121.
\item[] Schwartz, S.~H., 2007: Heat capacity, time constant, and sensitivity of Earth's climate system, \emph{J. Geophys. Res.}, {\bf 112}, D24S05.  
\item[] Schwartz, S.~H., 2008: Uncertainty in climate sensitivity: causes, consequences, challenges, \emph{Energy Environ. Sci.}, {\bf 1}, pp.~430-453.
\item[] Schwartz, S.~E., Charlson, R.~J., Kahn, R.~A., Ogren, J.~A.,  and Rodhe, H., 2010: Why Hasn't Earth Warmed as Much as Expected?\emph{ J. Climate}, \textbf{23}, 1453-2464.
\item[] Socolow, R.~H. and Lam, S.~H., 2007: Good enough tools for global warming policy making, \emph{Phil.~Trans.~R.~Soc.~A}, 
{\bf 365}, 1853, pp.~898-934.
\item[] Solomon, S, Qin, D., Manning, M., Chen, Z., Marquis, M., Averyt, K.~B., Tignor, M., Miller, H.~L., (eds). 2007: \emph{Climate Change 2007, (AR4): The Physical Science Basis} by WG I, and \emph{Mitigation of Climate Change} by WG III. Cambridge University Press. 
\item[] Stevens, B. and Bony, S., 2013: \emph{Physics Today}, pp.~29-34.
\item[] Stouffer, R.~J., 2004: Time Scales of Climate Response, \emph{J. Climate}, \textbf{17}, pp.209-217.
\item[] Weitzman, M., 2009: {On modeling and interpreting the economics of catastrophic climate change}, \emph{The Review of Economics
and Statistics}, {\bf 91}, No.~1, pp.~1-19.
\item[] Wigley, T.~M.~L., Jones, P.~D., and Raper, S.~C.~B., 1997: The observed global warming record: What does it tell us? 
\emph{PNAS}, {\bf 94}, No. 16, pp.~8314-8320.
\item[] Willis,~J.~K., Roemmich, D. and Cornuelle, B., 2004: Interannual variability in upper ocean heat content, temperature, and thermosteric expansion of global scales, \emph{J. Geophys. Res.}, {\bf 109}, C12036. 
\item[] Wijffels, S. E.,  Willis, J.,  Domingues, C. M.,  Barker, P., White, N. J., Gronell, A. , Ridgway, K. and  Church,  J. A., 2008: Changing eXpendable Bathythermograph fall-rates and their impact on estimates of thermosteric sea level rise. \emph{J. Climate}, {\bf 21}, pp.~5657-5672.
\end{description}

\renewcommand{\baselinestretch}{1}
\rm 

\begin{figure}[b]
\begin{center}
\includegraphics[width=1.1 \textwidth]{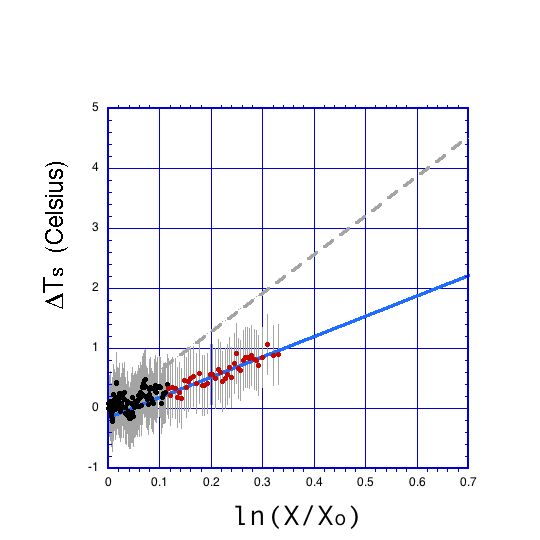}
\caption{$\Delta T_s (^\circ C)$ vs $\ln (X/X_o)$ (1850-2012). Temperature data from Brohan, Kennedy, Harris \emph{ \emph{\emph{et.~al.}}} (2006; HadCRUT3 dataset), Morice, Kennedy, Rayner \emph{et.~al.,} (2012)
and $CO_2$ data from Enting, Wigley and Heimann (1994; see its Fig.~B.1). 
Uncertainties in the individual $\Delta T_s (^\circ C)$ data are circa +/- 0.5$^\circ$C (Frank, 2010). 
Solid line is the least squares best-fit straight line ($\Td=2.3 ^\circ C; R=0.88$) for the post-1970 data ($\ln(X/X_o)>0.13$). The dashed line is \eqn{eqcs} plotted with $\Tcs=4.5 ^\circ C$---which is
the upper limit of the IPCC likely range. Note that $\ln 2\approx 0.7$.}
 \label{Hadley}
\end{center}
\end{figure}

\begin{figure}[h]
\begin{center}
\includegraphics[width=1.1 \textwidth]{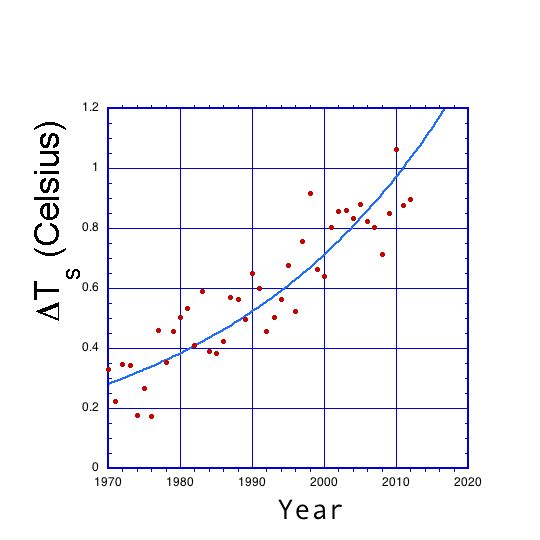}
\caption{$\Delta T_s (^\circ C)$ vs time  (Brohan  \emph{et al.}, 2006, Morice \emph{et.~al.}, 2012; post-1970 data only).
Solid line is the  least squares best-fit exponential line (R=0.90).  The time scale of the exponential line is $\tauT\approx 32$ years.  
Uncertainties in the individual $\Delta T_s (^\circ C)$ data are circa +/- 0.5$^\circ$C (Frank, 2010). }
 \label{Slope}
\end{center}
\end{figure}

\end{document}